\documentclass{elsart}

\usepackage[authoryear]{natbib}
\usepackage{graphicx}

\newcommand{\gapprox}{$\stackrel {>}{_{\sim}}$}   
\newcommand{\lapprox}{$\stackrel {<}{_{\sim}}$}

\begin{document}
\begin{frontmatter}
\title{{\it WISE} photometry of EXor sources and candidates}

\author[mp]{S.Antoniucci},
\ead{simone.antoniucci@oa-roma.inaf.it}
\author[mp]{T.Giannini},
\ead{teresa.giannini@oa-roma.inaf.it}
\author[mp]{D.Lorenzetti\corauthref{cor}}
\corauth[cor]{Corresponding author.}
\ead{dario.lorenzetti@oa-roma.inaf.it}
%

\address[mp]{INAF - Osservatorio Astronomico di Roma - Via Frascati, 33 -
00040 Monte Porzio Catone, Italy}

\begin{abstract}
 
We present a collection of {\it WISE} photometry of EXor sources and candidates (more recently identified). 
This represents the first complete survey of such objects in the mid-IR (3.4 - 22 $\mu$m) that was carried out 
with the same instrumentation. Two-color diagrams constructed with {\it WISE} data evidence a clear segregation 
between classical and newly identified sources, being these latter characterized by colder (and less evolved) 
circumstellar disks. By combining 2MASS and {\it WISE} data, we obtain spectral energy distributions (SED's) that
are compatible with the existence of an inner hole in the circumstellar disk. 
A compilation of all EXor observations given in the literature at 
wavelengths very similar to those of {\it WISE} is also provided. This allows us to study their mid-IR variability, 
which has been poorly investigated so far and without any coordination with shorter wavelengths surveys. The 
presented {\it WISE} photometry and the compilation of the literature data are intended as a first step
toward the construction of a significant database in this spectral regime. Preliminary indications on the mechanisms
responsible for the luminosity fluctuations are provided.

\end{abstract}


\end{frontmatter}

\section{Introduction}

Two decades ago a new class of young stellar objects (YSO's) was defined and studied by Herbig (1989, 2008),
who dubbed them as EXor sources, after the name of the prototype EX Lup. These pre-Main Sequence
stars are characterized by repetitive outbursts in optical light of amplitudes up to 4-5 mag.
Such outbursts typically last one year or less and are followed by long-lasting (5-10 yr) quiescence periods.
They are due to intermittent accretion events during which the matter accumulated in the inner parts of the 
circumstellar disks is violently transferred onto the central star inducing a shock on its surface, whose
cooling determines an impulsive increase of its luminosity. The physical mechanism and the observational
appearance of the EXor events make these objects substantially similar to FUors (Hartmann \& Kenyon 1985), although 
these latter present outbursts of larger intensity and remain in that state for longer period (tens of years). 

So far, about 20 EXor systems are known, which can be grouped into two sub-classes: the classical EXor
(Herbig 1989, 2008), and the newest identifications (listed in Lorenzetti et al. 2012a).
Historically, the first group was identified in the visual band, which means that these accretion-disk systems
are essentially unobscured (i.e. without significant envelopes). Later on, the increasing
availability of near-IR facilities quite naturally favoured the identification of
more embedded eruptive variables (second sub-class), typically associated with an optical-IR nebula.
The membership to this latter class often relies on the detection of sporadic
outbursts and does not stem from a more comprehensive analysis aimed at
checking all the properties typical of the classical
prototypes (i.e. repetitive outbursts, rapid brightening and slower fading, colors pre-and post-outburst).
The similarities and differences between the classes are discussed in more detail in Lorenzetti et al. (2012a). Here 
we recall that the newest ones are in general more embedded objects whose phenomenology is typically observed 
at infrared (IR) (1-10 $\mu$m) wavelengths. Indeed, given the mechanism that originates the outburst, there is 
no reason preventing the intermittent accretion from occurring also during a more embedded phase.
To study the modalities of the EXor brightness fluctuations, we are conducting since a few years a photometric 
(Lorenzetti et al. 2007) and spectroscopic (Lorenzetti et al. 2009, 2012b)
monitoring of EXors in the near-IR. 
Of particular interest is the mid-IR domain (3-25 $\mu$m), where many EXors emit most
of their energy. The spectral behaviour at these wavelengths is strictly related to disk and envelope regions 
located at radial distances (from the central star) where disk fragmentation and planet formation occur. The difficult access to facilities operating in this spectral range has hampered this kind of studies, apart from few important exceptions (K\'{o}sp\'{a}l et al. 2012 and references therein). However, during the last years IR surveys (e.g. {\it MSX, Spitzer, AKARI}) allowed the investigation of entire samples of sources at mid-IR frequencies. Very recently, the satellite {\it WISE} (Wright et al. 2010) has covered the whole sky at 3.4, 4.6, 12, and 22 $\mu$m, and we present here the {\it WISE} photometry of the EXors aiming at: ({\it i}) building up the first mid-IR catalog of these objects, which would serve as a reference for future and past (unfortunately very few and sparse) observations; ({\it ii}) providing 
a complete database for constructing spectral energy distributions; 
({\it iii}) fostering the study of the EXor variability in this still unexplored spectral range, practically unaffected by
extinction. The paper is organized in the following way: in Sect.2 we introduce the investigated EXors providing their {\it WISE} photometry, which is then discussed in Sect.3. In Sect.4 a database of the mid-IR observations 
collected so far in photometric band-passes very similar to those of {\it WISE} is given along with a short discussion 
on the detected variability cases. Finally, our concluding remarks are presented in Sect.5.

\section{Our sample and {\it WISE} photometry}

The sample we consider in the present paper is composed of the classical and newest EXors, which are listed in the upper 
and lower part of Table~\ref{wise:tab}, respectively. The {\it WISE}
photometry of the sources is given (columns 2 to 9) in terms of magnitudes and relative errors. The
physical parameters of all these objects (with the corresponding
references) can be found in Table 1 of Lorenzetti et al. 2012a.

As described by Cutri et al. 2012,
all magnitudes listed in the WISE Source Catalog have been photometrically calibrated using observations of a network 
of standard stars that are located near the ecliptic polar caps. In that paper, the process by which 
instrumental source magnitudes were converted to calibrated magnitudes is also described. All the sources listed in Table~\ref{wise:tab}
present a photometric quality flag AAAA, which means that the source is detected in all 4 bands with a signal-to-noise 
ratio $>$ 10. Saturation begins to affect sources brighter than approximately 8.1, 6.7, 3.8, and -0.4 mag at 3.4, 4.6, 12, and 22 $\mu$m, respectively.  
{\it WISE} profile-fit photometry of saturated sources is obtained by fitting the Point Spread Function (PSF) to the non-saturated pixels in the profile wings. 
The accuracy of photometric measurements of saturated sources degrades with increasing brightness. For our sources, however, saturation typically affects less than 30\% of the pixels, hence this effect does not alter our conclusions, as discussed below in the specific sections. 
Column 9 gives the variability flag, whose definition and significance is discussed in the next paragraph.
As additional information, columns 10 and 11 of Table~\ref{wise:tab} report the distance (in arcsec) 
between the positions of the {\it WISE} source and associated 2MASS PSC object and the Position Angle (in degrees from
N to E) of the relative separation vector. In the large majority of cases (14 out of 18)
such distance is \lapprox~1 arcsec, which proves that 2MASS and {\it WISE} are almost certainly looking at the same source.


{\it WISE} also provides a record of multi-epoch (on a daily time-scale) measurements from the individual images that were 
co-added to extract the Catalog fluxes given in Table~\ref{wise:tab}. Typically, a dozen of independent exposures were acquired for each point near the ecliptic so that these multiple observations could be fruitfully investigated to look for
clues of short time-scale variability. The{\it WISE} Catalog entries include a variability flag 
(see Table~\ref{wise:tab} - col.9)
consisting of a four-character string (one character per band) giving a measure of the probability that the source is variable in each band. The derivation of this flag is discussed by Cutri et al. 2012 and by Hoffman et al. 2012; both descriptions emphasize the need to exclude {\it WISE} single-exposure measurements that \textit{i}) come from low quality images, \textit{ii}) are saturated, \textit{iii}) have large reduced $\chi^2$ values in profile fitting, plus other reasons.
Hence, indication of variability from the Single-exposure Catalog can be conservatively taken into account only for those bands that present a variability flag value \gapprox~7 (see Table~\ref{wise:tab} caption). In our case, eight sources (namely VY Tau, DR Tau, V1118 Ori, NY Ori, LDN1415 IRS, V2775 Ori, V2492 Cyg, and V2493 Cyg) have a high probability of presenting short-time variability in one or more bands.
In the following we will cautiously use these data as a further support to cases of (independently ascertained)
daily variability.\ 
Finally, another {\it WISE} database is in principle available for studying the variability of the objects, namely the preliminary release of Post-Cryogenic data. However, some caution should be exercised when comparing cryo and post-cryo photometry, because there is only a preliminary calibration available at the moment. For this reason we decided not to include these data in the present paper.

\section{Results and discussion}

\subsection{{\it WISE} color diagrams}

\begin{figure}
\includegraphics[width=9.5cm]{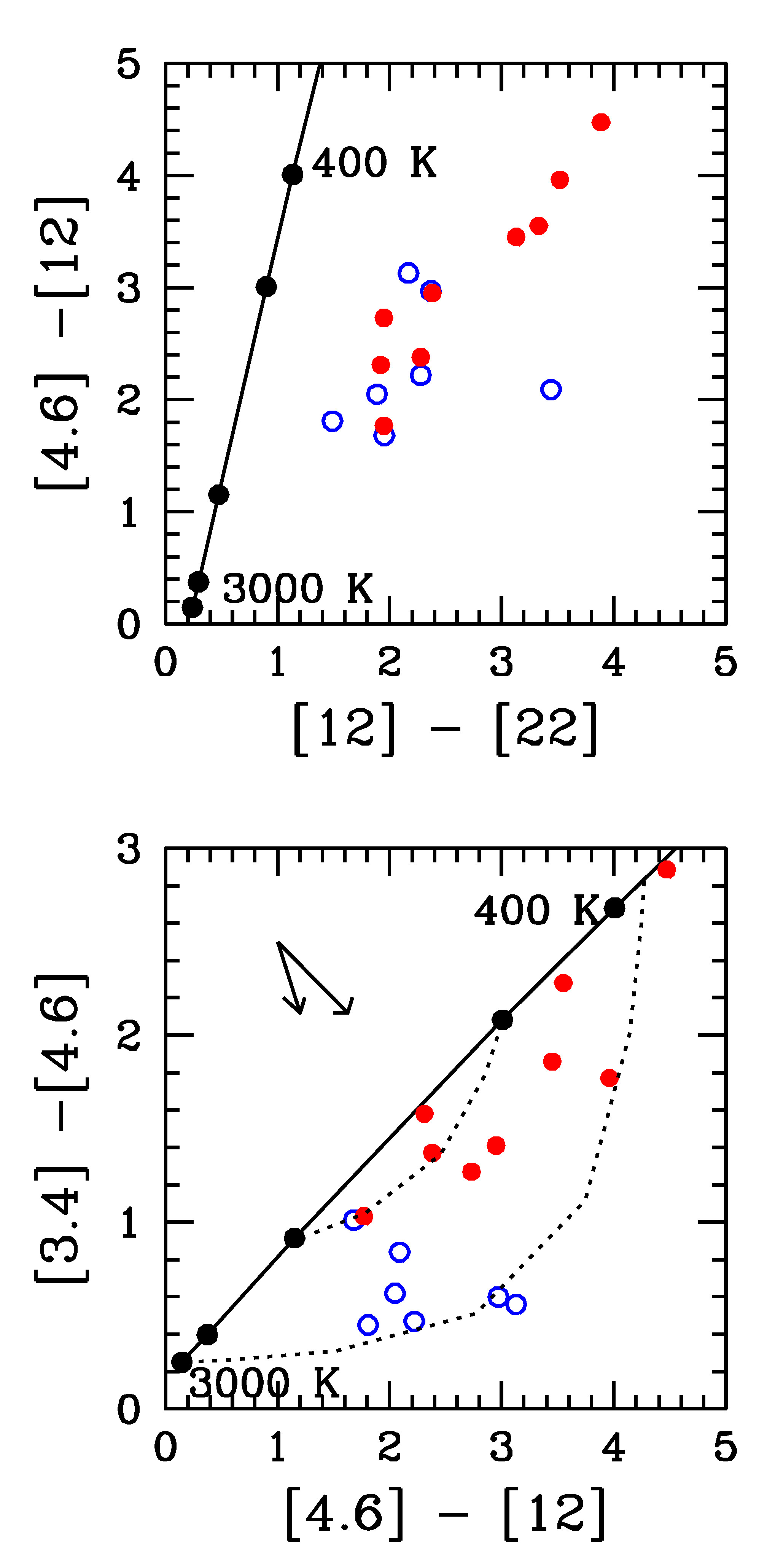}
\caption{\label{fig1:fig} {\it WISE} two color diagrams of EXors. The colors of the classical (newest) EXors are depicted with blue (red) open (solid) symbols. 
The size of the error bars is typically comparable to the size of the points.
The black-body colors (at different temperatures) are indicated by the black straight line, with the solid circle marks corresponding to a temperature of 400, 500, 1000, 2000, and 3000 K, from right to left). 
The black dashed lines in the lower panel indicate the colors of the sum of two black-bodies, where the 
surface of coldest one (in the range 380-500 K) is progressively increased with respect to that of the warmest (in the range 1000-3000 K). The reddening vectors correspond to an extinction A$_V$ = 10 mag and were computed from Rieke \& Lebofsky (1985); two different values for the 12$\mu$m extinction were assumed: equal to A$_{\lambda}$ at 10.5$\mu$m in Rieke \& Lebofsky table (leftmost vector) and equal to A$_{\lambda}$ at 12.0$\mu$m .}
\end{figure}

The {\it WISE} photometry given in Table~\ref{wise:tab} is presented in Figure~\ref{fig1:fig} in form of two color
plots  [3.4-4.6] vs. [4.6-12] (lower panel) and [4.6-12] vs. [12-22] (upper panel). Possible saturation (only for the 
brightest objects) marginally affects the {\it WISE} photometry in the four bands, thus its effect on colors is expected 
to be negligible. EXors are located in loci
framed by a color dispersion \gapprox~2 mag: between 0.4 and 3 mag for the [3.4-4.6] color and between 1 and 3 mag for 
all the other colors. Expectedly, these are the loci typically pertaining to the pre-Main Sequence objects characterized
by a significant level of IR excess over the photospheric continuum. Classical (newest) EXors are represented in
Figure~\ref{fig1:fig} with blue open (red solid) symbols, respectively. A segregation between the two sub-classes is 
noticeable in all the colors, being the newest also the reddest objects. Particularly evident is the 
segregation in the [3.4-4.6] color, which is $>$ 1 for the newest and $<$ 1 for the classical EXors,
the unity value corresponding to a ratio F$_{\nu}$(3.4$\mu$m)/F$_{\nu}$(4.6$\mu$m) $\approx$~4. Although classical objects
are certainly less obscured than the newest ones, the color separation between the two classes is certainly due also to intrinsic effects.
Indeed, it is impossible to superpose newest EXors to the classical ones, not even considering a differential extinction larger than 10 mag (arrows depicted in Figure~\ref{fig1:fig}). Hence, the SED's of sources of the two sub-classes should be intrinsically different (see Sect.3.2). 
From Figure~\ref{fig1:fig} it is also evident that the {\it WISE} colors of EXor objects are not consistent with a single black-body, but
can be reproduced considering different dust components with the highest temperature ranging between 1000 and 3000 K and the lowest one
between 380 and 500 K. Oversimplifying, in Figure~\ref{fig1:fig} we depict only pairs of black-body functions, but a more realistic stratification of 
temperatures should be envisaged.

\subsection{EXor near- and mid-IR SED's}

\begin{figure}
\includegraphics[width=16cm]{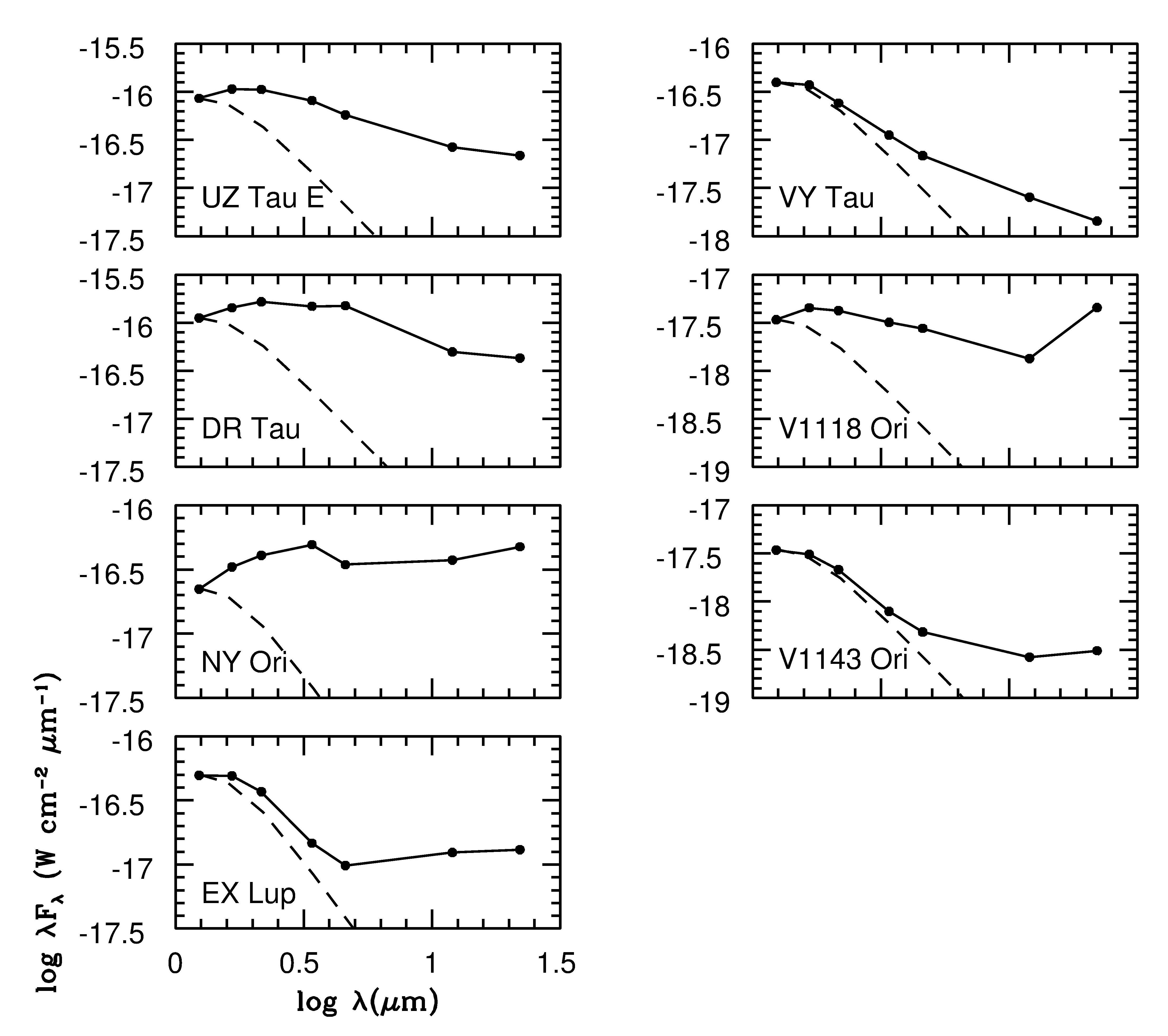}
\caption{\label{fig2:fig} SED's of the {\it classical} EXor sources in the range 1.25-22 $\mu$m, composed by 2MASS and {\it WISE} fluxes. Dashed line represents a median photosphere of stars in the spectral type range K5-M5 and normalized to the J-band flux of the source.}
\end{figure}

\begin{figure}
\includegraphics[width=16cm]{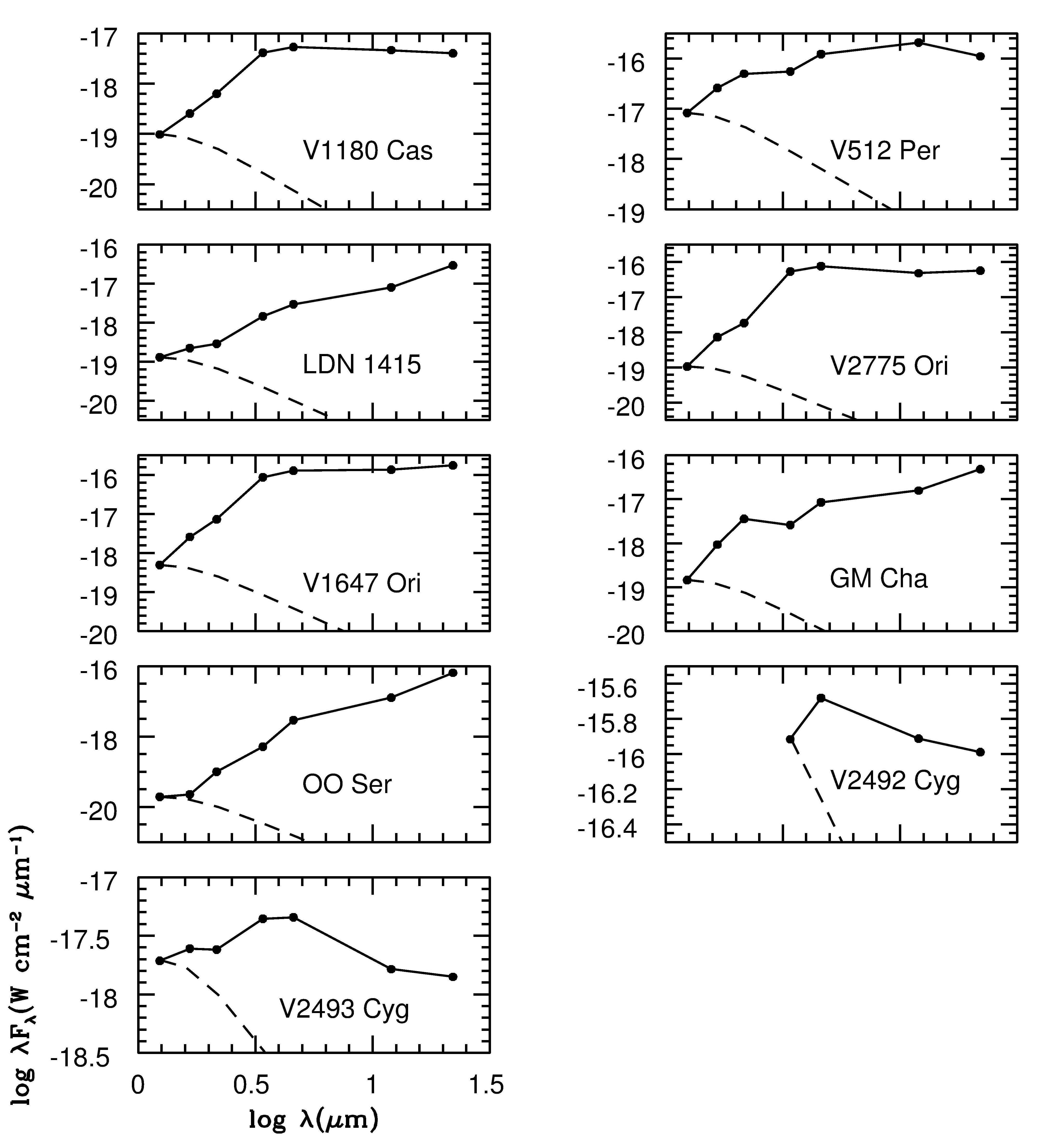}
\caption{\label{fig3:fig} Same as Figure~\ref{fig2:fig} for the {\it newest} EXor sources.}
\end{figure}

We obtained the near- and mid-IR SED's of EXors by identifying the counterparts at shorter wavelengths (JHK bands) of the
sources listed in Table~\ref{wise:tab}. Unfortunately, a single epoch SED cannot be constructed since coordinated observations in different spectral ranges are very rare (see Sect. 4). However,  to rely on a data-base as uniform as possible, we considered the 2MASS (Strutskie  
et al. 2006) counterparts that match our sources within 2 arcsec or less (see Table~\ref{wise:tab}). All the listed EXors (except one, V2492 Cyg) have been associated to a 2MASS counterpart and the resulting SED's are depicted in Figures~\ref{fig2:fig} and \ref{fig3:fig}. For comparison, a median stellar photosphere in the spectral range K5-M5 is 
also plotted in each panel (Hern\'{a}ndez et al. 2007), normalized to the J-band flux of the source. 
As anticipated by the two-color diagrams, also the SED shapes of the two sub-classes are substantially different. Emission from all the classical EXors (except for NY Ori) declines with increasing wavelength for $\lambda$ \lapprox~10 $\mu$m; at $\lambda$ 
\gapprox~10 $\mu$m their SED's tend to increase, showing an excess more pronounced than that at shorter wavelengths.
Conversely, the newest candidates (apart from two exceptions) have SED's that monotonically increase with wavelength. 
The spectral distributions of both classes are fully consistent with those predicted by D'Alessio et al. (1999), namely they originate from a temperature stratification related to a more (classical EXor) or less (newest ones) evolved circumstellar disk.
To give a quantitative evaluation of the EXor SED's we introduce as empirical indicator the parameter $\varepsilon$, i.e. the ratio 
of the total 1-22 $\mu$m luminosity to the luminosity of a median K5-M5 photosphere. In this first-order approximation the
J-band flux is assumed to be entirely photospheric, as implied by the normalization used to show the SED's in Figures~\ref{fig2:fig} and \ref{fig3:fig}. The obtained values of $\varepsilon$ are given in the last column of Table~\ref{wise:tab}: the relatively modest excess presented by the classical EXors 
is accounted for by their $\varepsilon$ values \lapprox~20, while the newest ones present $\varepsilon$ typically $\gg$~200, except for the
two unobscured T Tauri stars V2492 and V2493 Cyg.
This short analysis evidences how the mid-IR emission is crucial for observationally defining the EXor class of objects. Finally, and more speculatively, we can observe that the majority of EXor SED's (10 out of the 16 depicted in 
Figures~\ref{fig2:fig} and \ref{fig3:fig}) seem to show a deficit of emission around 3 $\mu$m. This appearance should be considered very cautiously since that wavelength is exactly the matching point between 2MASS and {\it WISE} photometry, and 2MASS data were obtained about ten years before the {\it WISE} advent. On the other hand, a systematic trend according to which {\it WISE} data would have been taken when all the sources were in a stage more quiescent than 2MASS, sounds remarkably strange. 
Saturation could cause the aperture photometry to be systematically underestimated, so that, accounting for this effect, the observed deficit could be partially reduced in DR Tau and V512 Per. However, the same effect would definitely enhance the discontinuity in NY Ori, V1647 Ori, and V2775 Ori. In any case, if such an emission deficit will be confirmed by forthcoming observations, it reasonably indicates a lack of emission in that spectral range, which is usually interpreted in 
terms of a circumstellar disk with an inner hole, a scenario which is largely supported by both photometric (Sipos et al. 2009, Lorenzetti et al 2012a) and mainly interferometric (Akeson et al. 2005, Eisner et al. 2010) studies.

\section{Mid-IR variability}

\begin{figure}
\includegraphics[width=16cm]{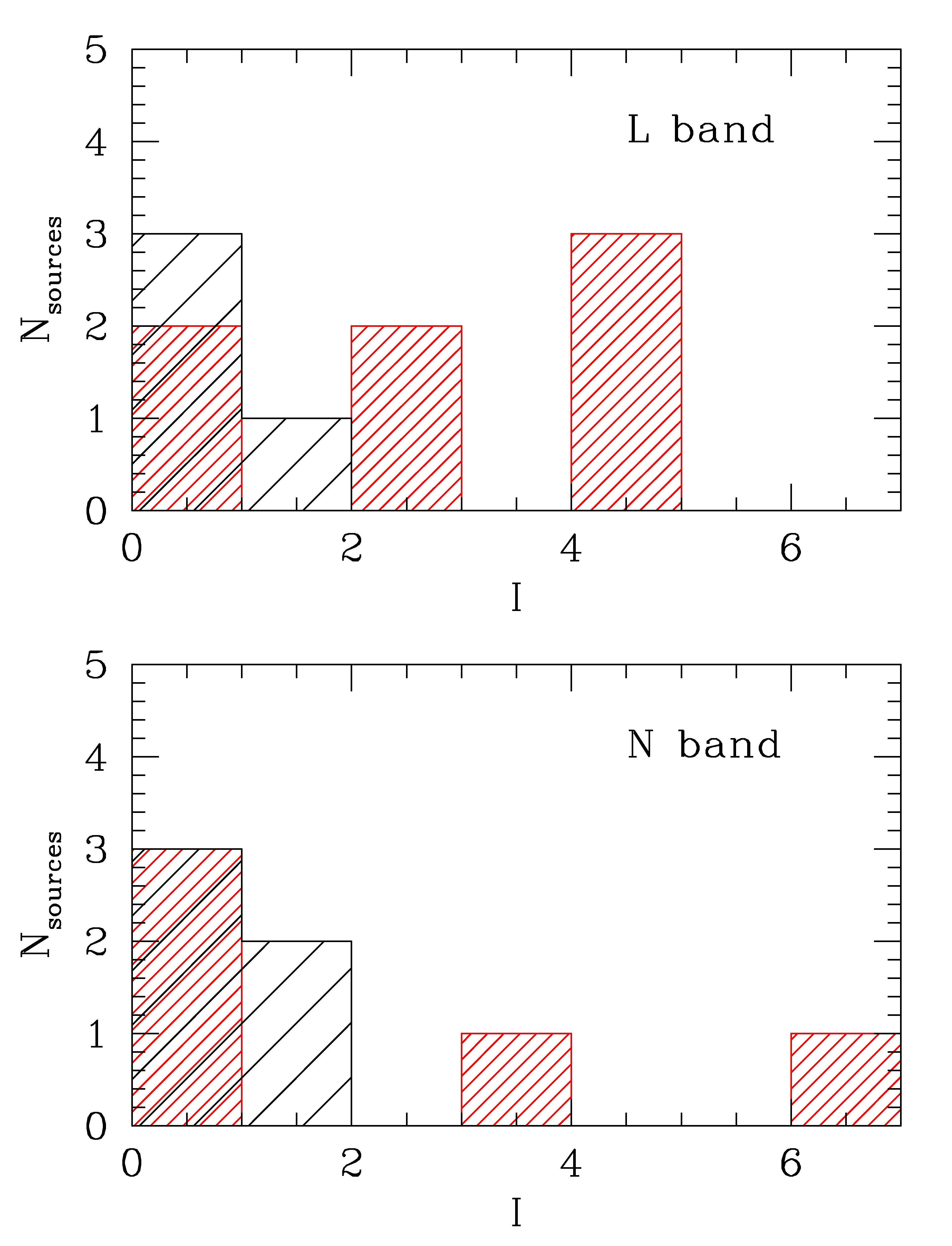}
\caption{\label{fig4:fig} Distribution of the EXor fluctuations expressed in terms of the variability index \textit{I} (see text) in the L and N bands (see Tables~\ref{variabLM:tab} and \ref{variabNQ:tab}).}
\end{figure}

To ascertain if EXors present significant fluctuations in the {\it WISE} mid-IR range as they certainly do in the optical/near-IR, we have collected in Tables~\ref{variabLM:tab} and \ref{variabNQ:tab} the observations carried out in the L 
($\sim$3$\mu$m), M ($\sim$5$\mu$m), N ($\sim$10$\mu$m), and Q ($\sim$20$\mu$m) bands and presented so far in the literature.
To construct these Tables we have adopted the following data: all the 
ground-based observations, IRAS 12 and 25 $\mu$m, MSX 12 and 21 $\mu$m, {\it Spitzer} IRAC 3.4  and 4.5 $\mu$m, and 
{\it Spitzer} MIPS 24 $\mu$m. Data taken in different, although adjacent, bands (eg. IRAC 5.8 and 8.0 $\mu$m) have not been considered because of the large difference between their and {\it WISE} band-passes. The epoch of any single observation and the effective wavelength of the used band-pass are also given in Tables~\ref{variabLM:tab} and \ref{variabNQ:tab}. 
For an easier comparison, the values corresponding to the L, M (N, Q) bands are given in magnitudes (in Jansky): in some cases, the photometric values found in the literature have been converted into magnitude or fluxes by adopting the zero-magnitude fluxes given in the WEB resources of the Gemini Observatory (http://www.gemini.edu/?q=node/11119), whenever the conversion values are not given in the original paper. This procedure, together with the comparison between band-passes which are not perfectly coincident, is affected by instrumental and calibration effects and may cause a scatter 
that we conservatively estimate up to $\pm$0.5mag.
This scatter is much larger than the errors of any single measurement, nevertheless it has to be
considered as the real uncertainty. Hence, fluctuations up to that amplitude have to be considered negligible for the present work, unless they have been detected with the same instrumentation: in these latter cases the fluctuations are assumed to be significant if their amplitude is larger
than the 3$\sigma$ uncertainty. IRAS fluxes are
listed just for completeness, but not considered for the quantitative comparison, because they are taken with a beam size much larger than that of other observations. Data in Tables~\ref{variabLM:tab} and \ref{variabNQ:tab}, complemented with {\it WISE} observations, are summarized in Figure~\ref{fig4:fig}, which shows a histogram of sources as a function of the amplitude of their fluctuations in the two most illustrative bands (L and N), for both classical (dashed black line) and newest (solid red line) EXors. Only the long-term (months, years) variations are considered here; the measurements dealing with fluctuations on a daily time-scale will be discussed below in this Section. 
We have computed (for any given source) the averaged values $\overline{L}$ and $\overline{F_N}$ of the L-band magnitude and N-band flux, respectively. Then,
$\Delta$mag values have been computed for any single observation as 
$\mid L - \overline{L} \mid$ and $\mid 2.5 \cdot Log(F_N/\overline{F_N}) \mid$. Therefore, for each source we have as many $\Delta$mag values
as the available observations; in Figure~\ref{fig4:fig} we depict the maximum value of the variability index $I$, defined as $\Delta$mag/0.5 mag or
$\Delta$mag/3$\sigma$, depending on how the observational set was obtained (i.e. different or same instrumentation, respectively).

As mentioned before, the fluctuations corresponding to the 
first bin are conservatively neglected; larger $I$ values are
mainly found for the newest objects: this fact is due in part to their
intrinsically different nature (i.e. dominated by colder dust), but it is also a bias related to the larger
number of mid-IR observations available for the newest sources. The amplitude of these fluctuations is comparable with
that occurring in the near-IR, where a decreasing trend with increasing wavelength was noted (Lorenzetti et al. 2007).
In other words, it seems that from $\lambda$ $>$ 2-3 $\mu$m the amplitude of the fluctuations becomes independent on $\lambda$.
This likely means that the decreasing trend of $\Delta$mag in the optical and near-IR bands is essentially related to the stellar component only,
while in the frequency regime where dust emission starts to dominate, any spectral trend is not that evident.\


The time-scale of the mid-IR variability is even more important than the variability itself, since it can elucidate on the
physical mechanism responsible for the fluctuations. The aforementioned analysis is typically based on time-scales of months, years or even
longer, while only investigations at shorter time-scales can effectively constrain the working physical mechanism(s).
To that scope, it is worthwhile to extract from Tables~\ref{variabLM:tab} and \ref{variabNQ:tab} the observations 
executed on shorter time-scales (days) and with the same instrumentation, so as to avoid uncertainties coming from possible inter-calibration issues.
These cases are summarized in Table~\ref{timescale:tab} where we list epoch and
spectral band of the monitoring (cols.2 and 3), the number of observations available in that period (col.4), the parameter 
$\mid x - \overline{x} \mid_{max}$ (col.5), which corresponds to the largest variation 
(with respect to the mean value) observed in that period, and the relative 3$\sigma$ uncertainty (col.6).
This result is compatible with the {\it WISE} multi-epoch data discussed in Sect.2: indication of short-time variability 
was given there for the source DR Tau, while V1180 Cas, and V512 Per did not present enough data to reach a firm conclusion.\

These very short time-scales can be related to the response time of the dust particles to the energy input provided by the accretion event or, alternatively, to fast changes of the dust distributions (Kun et al. 2011). Basing upon a still rather poor statistics (3 positive cases out of 3 investigated), we can tentatively conclude that all EXors could be affected by variations of the same amplitude on
the same time-scale (days): if this were the case, the response of the dust particle should be preferred to the 
dust re-arrangements, since the constancy of the observables seems more compatible with an intrinsic property of the dust grains, more than with a changing of the dust morphology, which should unexpectedly follow the same modalities. 
However, to ascertain the role of the two possible mechanisms, (quasi-)simultaneous ground-based observations in the near- and mid-IR (L, M bands) should be in order.

\section{Concluding remarks}

We have compiled the first collection of {\it WISE} photometry (3.4 - 22 $\mu$m) of the mid-IR emission of
both classical and newest EXor objects. Analysing the presented data the
following conclusions can be summarized:

\begin{itemize}
\item The {\it WISE} photometry along with the compilation of all the mid-observations
retrieved from the literature represents the first step for constructing a mid-IR database 
of the EXor continuum emission.
\item In the {\it WISE} two-color plots the classical and the newly defined EXors are
distinguishable since the latter present a more prominent emission by their circumstellar
disks, which appears also colder and less evolved.
\item EXor SED's confirm the previous result providing piece of evidence in favour of the presence
an inner hole in the circumstellar disk, as predicted by theory and observed 
by interferometric investigations.
\item We began a search for mid-IR variability and preliminary results indicate that
({\it i}) the amplitude of the fluctuations is comparable to that occurring in near-IR without any decreasing trend
with wavelength and that ({\it ii}) all the sources sampled on a daily time-scale present significant variations. 
Implications on possible mechanisms responsible for this behaviour have been discussed.
\end{itemize}

\clearpage

\section{Acknowledgements}

This publication makes use of data products from the Wide-field Infrared Survey Explorer, which is a joint project of the University of California, Los Angeles, and the Jet Propulsion Laboratory/California Institute of Technology, funded by the National Aeronautics and Space Administration. 
It makes also use of data products from the Two Micron All Sky Survey, which is a joint project of the University of Massachusetts and the Infrared Processing and Analysis Center/California Institute of Technology, funded by the National Aeronautics and Space Administration and the National Science Foundation.

\clearpage


\begin{table}
\begin{center} 

\caption{WISE photometry of EXor objects. \label{wise:tab}} 
\medskip
{\scriptsize
\begin{tabular}{lcccccccccccc}
\hline
\medskip
Source&3.4$\mu$m&err 1&4.6$\mu$m&err 2&12$\mu$m&err 3&22$\mu$m&err 4&var.flag$^a$&\multicolumn{2}{c}{(2MASS)}&$\varepsilon$\\
                  &                    \multicolumn{8}{c}{(mag)}             &  &  D (") &  PA (deg) \\
\hline

UZ Tau E          & 6.34    & 0.04  & 5.72   & 0.02  & 3.67   & 0.01  & 1.78   & 0.01  & 00nn & 0.3       & 223.8 &  5.8\\
VY Tau            & 8.48    & 0.02  & 8.03   & 0.02  & 6.22   & 0.02  & 4.73   & 0.03  & 0019 & 0.1       & 150.3 &  1.7\\
DR Tau            & 5.68    & 0.06  & 4.67   & 0.04  & 2.99   & 0.01  & 1.04   & 0.02  & 0085 & 0.1       &  81.0 &  8.9\\
V1118 Ori         & 9.85    & 0.02  & 9.01   & 0.02  & 6.92   & 0.10  & 3.48   & 0.05  & 0019 & 0.1       & 150.3 &  9.7\\
NY Ori            & 6.87    & 0.03  & 6.27   & 0.02  & 3.30   & 0.02  & 0.93   & 0.03  & 009n & 0.4       & 275.0 & 21.6\\
V1143 Ori         & 11.37   & 0.02  & 10.90  & 0.02  & 8.68   & 0.03  & 6.40   & 0.06  & 0011 & 0.2       & 111.0 &  1.8\\
EX Lup            & 8.19    & 0.02  &  7.63  & 0.02  & 4.50   & 0.01  & 2.33   & 0.02  & 1110 & 0.2       & 195.9 &  3.5\\
%
%
\hline
V1180 Cas         & 9.56    & 0.02  &  8.29  & 0.02  & 5.56   & 0.01  &  3.61  & 0.04  & nnnn & 0.18      & 258.4 & 534\\
V512 Per          & 6.75    & 0.03  &  4.89  & 0.03  & 1.44   & 0.01  & -1.69  & 0.01  & 10nn & 0.28      & 116.1 & 212\\
LDN1415 IRS       & 10.71   & 0.02  &  8.94  & 0.02  & 4.94   & 0.01  &  1.46  & 0.01  & 7941 & 4.6       &  5.8  & 1041\\
V2775 Ori         & 6.78    & 0.03  &  5.41  & 0.03  & 3.03   & 0.01  &  0.75  & 0.01  & 0019 & 0.10      &  23.2 & 5951\\

V1647 Ori         & 6.26    & 0.05  &  4.85  & 0.03  & 1.90   & 0.01  & -0.48   & 0.01 & 0151 & 0.10      & 241.6 & 3233\\
GM Cha            & 10.07   & 0.03  &  7.79  & 0.02  & 4.24   & 0.01  &  0.91   & 0.01 & 1642 & 0.30      &  40.8 & 1706\\
OO Ser            & 11.83   & 0.03  &  8.94  & 0.02  & 4.47   & 0.01  &  0.59   & 0.01 & 0320 & 0.70      & 192.0 & $>$1.3 10$^4$\\
V2492 Cyg         &  5.90   & 0.05  &  4.32  & 0.03  & 2.01   & 0.01  &  0.09   & 0.01 & 1097 & ---       & ---   &  5.5\\
V2493 Cyg         &  9.50   & 0.02  &  8.47  & 0.02  & 6.70   & 0.02  &  4.75   & 0.05 & 9999 & 2.53      & 196.0 & 14.9\\
%
%
\hline\hline

\end{tabular}}
\end{center}
\medskip
- The fluxes corresponding to zero magnitude are: 309.5, 171.8, 31.67, and 8.363 Jy for the bands at 3.4, 4.6,
12, and 22~$\mu$m, respectively (Cutri et al. 2012). \\
- The quality flag is AAAA for all sources.\\
- Saturation affects photometry for sources brighter than approximately 8.1, 6.7, 3.8, and -0.4 mag at 3.4, 4.6, 12, and 22 $\mu$m, respectively.\\ 

$^a$ - {\it n} indicates insufficient or inadequate data to make a variability determination. Values of 0 through 5 can generally be regarded as non-variable sources in that band. Values of 6 and 7 can be regarded as potentially variable with small amplitudes. Objects with a value of 8 or 9 are most likely flux variables in the given band.
\bigskip

\end{table} 

\clearpage

\scriptsize
\begin{table}
\begin{center} 

\caption{L and M band photometry of EXors. \label{variabLM:tab}} 
\medskip
{\scriptsize
\begin{tabular}{l|c|cc|cc}
\hline
\medskip
Source    &  Epoch    &        \multicolumn{2}{c}{L band}               &       \multicolumn{2}{c}{M band}         \\ 
          &           &  (mag)          & [$\lambda_{eff}$] - Ref       &    (mag)       &  [$\lambda_{eff}$] - Ref\\                   
\hline
UZ Tau E  &  73/74    &  6.2 $\pm$ 0.2  &  [3.6] -    1                 &                 &                       \\
          &  Nov 76   &  6.29           &  [3.4] -    2                 &                 &                       \\
          &  Dec 76   &  6.4 $\pm$ 0.1  &  [3.4] -    2                 & 5.6 $\pm$ 0.3   &  [4.8] - 2            \\
          &  Nov 77   & 6.04 $\pm$ 0.02 &  [3.5] -    3                 &                 &                       \\
          &  Jan 81   & 6.29            &  [3.4] -    4                 &  5.55           &  [4.8] - 4            \\
          & Dec 01$^a$& 6.42 $\pm$ 0.17 &  [3.5] -    5                 &                 &                       \\  
          &  Mar 04   & 5.98 $\pm$ 0.03 & IRAC[3.4] - 6                 & 5.46 $\pm$ 0.04 &  IRAC[4.5] - 6        \\ 
\hline
VY Tau    &  73/74    &  8.6 $\pm$ 0.2  &  [3.6] -    1                 &                 &                       \\
          &  Dec 73   &  8.3 $\pm$ 0.1  &  [3.5] -    7                 &                 &                       \\        
          &  Nov 77   &  8.9 $\pm$ 0.1  &  [3.5] -    3                 &                 &                       \\
          &  Dec 81   & 8.50 $\pm$ 0.06 &  [3.5] -    8                 &                 &                       \\
          &  Dec 01   & 8.31 $\pm$ 0.09 &  [3.5] -    5                 &                 &                       \\
\hline       
DR Tau    &  72/74    &  5.9 $\pm$ 0.2  &  [3.6] -    1                 &                 &                       \\
          &  Nov 73   & 5.8 $\pm$ 0.05  &  [3.5] -    7                 &  5.0 $\pm$ 0.2  &  [4.8] - 7            \\ 
          &  Nov 77   & 4.83$\pm$ 0.05  &  [3.5] -    3                 &                 &                       \\
          &  Dec 81   & 5.34$\pm$ 0.04  &  [3.5] -    8                 &                 &                       \\
          &  Dec 81   & 5.15$\pm$ 0.04  &  [3.5] -    8                 & 4.17            &  [4.8] - 4            \\
          &  Nov 87   & 5.01 - 5.32     &  [3.8] -    9                 & 4.47 - 4.69     &  [4.7] - 9            \\
          &  Sep 88   & 4.93 - 5.46     &  [3.8] -    9                 &                 &                       \\ 
          &  96/98    & 5.3$^b$         &  [3.4] -    10                &                 &                       \\ 
          &  Mar 04   &                 &                               & 4.5             &  IRAC[4.5] - 6        \\ 
\hline
V1118 Ori &  Oct 04   &9.85 $\pm$ 0.002 &  IRAC[3.4] - 11               & 9.08$\pm$ 0.002 &  IRAC[4.5] - 11       \\       
\hline
NY Ori    &  Sep 75   &  $>$ 9          &  [3.6] -  12                  &                 &                       \\        
\hline
V1143 Ori &           &                 &                               &                 &                       \\       
\hline
EX Lup    &  Jun 73   &  $>$ 8.7        &   [3.4] -  13                 &                 &                       \\
          &  Apr 82   & 8.02$\pm$ 0.01  &   [3.8] -  14                   &                 &                     \\  
          &  Apr 82   & 8.05$\pm$ 0.01  &   [3.4] - 15                  & 7.54$\pm$ 0.05  &  [4.8] - 15           \\  
          &Feb/Sep 97 &(8.0-8.2)$\pm$0.1&   [3.6] - 16                  &                 &                       \\  
          &  Mar 05   & 7.92 $\pm$ 0.02 &  IRAC[3.4] - 16               & 7.35$\pm$ 0.02  &  IRAC[4.5] - 16       \\   
\hline
\end{tabular}}
\end{center}
\end{table} 
\addtocounter{table}{-1}

\begin{table}
\begin{center} 
\caption{Continued. \label{variabLM:tab}} 
\medskip
{\scriptsize
\begin{tabular}{l|c|cc|cc}
\hline
\medskip
Source    &  Epoch    &        \multicolumn{2}{c}{L band}               &       \multicolumn{2}{c}{M band}         \\ 
          &           &  (mag)          & [$\lambda_{eff}$] - Ref       &    (mag)       &  [$\lambda_{eff}$] - Ref\\                   
\hline

V1180 Cas &  Mar 09   & 9.50$\pm$ 0.06  &   IRAC[3.4] - 17              & 8.58$\pm$ 0.04  &  IRAC[4.5] - 17       \\  
          & Oct/Nov 09&(8.99-9.28)$\pm$0.01  &   IRAC[3.4] - 17         &(8.05-8.30)$\pm$ 0.01  &  IRAC[4.5] - 17 \\      
\hline
V512 Per  &  $<$ 83   &6.37$\pm$0.01    & [3.5] - 18                    & 4.71$\pm$0.09   &   [4.6] - 18          \\
          &  Sep 88   &(5.20-5.59)$\pm$0.03  & [3.91] - 19              &    $>$ 4.2      &   [4.7] - 19          \\
          &  Jan 89   &(5.3-5.7)$\pm$0.2     & [3.91] - 19              &    $>$ 4.7      &   [4.7] - 19          \\ 
          &  Feb 90   & 5.3$\pm$0.1          & [3.91] - 19              &    $>$ 4.4      &   [4.7] - 19          \\   
          &  Oct 90   & 5.28$\pm$0.02        & [3.91] - 19              & 4.2$\pm$0.1     &   [4.7] - 19          \\  
          &  Sep 91   & 5.88$\pm$0.05        & [3.6] - 20               &                 &                       \\
          &  Mar 92   & 6.11$\pm$0.05        & [3.6] - 20               &                 &                       \\ 
          &  Nov 92   & 6.11$\pm$0.05        & [3.42] - 20              &                 &                       \\ 
          &  Feb 93   & 5.96$\pm$0.05        & [3.42] - 20              &                 &                       \\ 
          &  Oct 93   & 6.27$\pm$0.05        & [3.42] - 20              &                 &                       \\ 
          &  Dec 93   & 6.48$\pm$0.05        & [3.42] - 20              &                 &                       \\ 
          &  Dec 93   & 6.60$\pm$0.05        & [3.42] - 21              &                 &                       \\
          &  Sep 04   & 7.45$\pm$0.07        & IRAC[3.4] - 22           & 6.39$\pm$0.09   &  IRAC[4.5] - 22       \\
\hline
LDN1415IRS&           &                      &                          &                 &                       \\       
\hline
V2775 Ori &Feb/Oct 04 & 8.46$\pm$0.05        & IRAC[3.4] - 23           & 7.66$\pm$0.05   &  IRAC[4.5] - 23       \\
\hline
V1647 Ori &  Mar 04   & 5.33                 & IRAC[3.4] - 24           & 4.41            &  IRAC[4.5] - 24       \\
          &  Feb 07   & 7.6$\pm$0.1          & [3.4] - 25               &                 &                       \\
          &  Sep 08   & 5.8$\pm$0.1          & [3.4] - 25               &                 &                       \\ 
          &  Feb 11   & 5.6$\pm$0.1          & [3.4] - 25               &                 &                       \\         
\hline
GM Cha    &  Mar 99   &8.32$\pm$0.05         & [3.5] - 26               &                 &                       \\
          &  Jul 04   &8.42$\pm$0.08         & IRAC[3.4] - 27           & 6.98$\pm$0.07   &  IRAC[4.5] - 27       \\        
\hline
OO Ser    &  Oct 95   &   6.9                & [3.8] - 28               &  5.2            &  [4.8] - 28           \\
          &  Feb 96   &   8.1$\pm$0.1        & [3.6] - 29               &  5.36$\pm$0.01  &  [4.8] - 29           \\ 
          &  Apr 96   &   8.2$\pm$0.1        & [3.6] - 29               &  5.45$\pm$0.03  &  [4.8] - 29           \\
          &  Sep 96   &   8.22$\pm$0.08      & [3.6] - 29               &  5.84$\pm$0.04  &  [4.8] - 29           \\   
          &  Oct 96   &   8.9$\pm$0.4        & [3.6] - 29               &  5.9$\pm$0.4    &  [4.8] - 29           \\ 
          &  Mar 97   &   8.9$\pm$0.1        & [3.6] - 29               &  6.13$\pm$0.05  &  [4.8] - 29           \\  
          &  Apr 97   &   9.06$\pm$0.08      & [3.6] - 29               &  5.98$\pm$0.04  &  [4.8] - 29           \\ 
          &  Sep 97   &                      &                          &  6.3$\pm$0.4    &  [4.8] - 29           \\
          &  Apr 04   & 11.5$\pm$0.1         & IRAC[3.4] - 29           & 8.35$\pm$0.02   &  IRAC[4.5] - 29       \\ 
\hline
V2492 Cyg &  Jul 06   &   8.94$\pm$0.09      & IRAC[3.4] - 30           &  7.6$\pm$0.1   &  IRAC[4.5] - 30        \\
          &  Sep 10   &   5.4$\pm$0.1        &  [3.8] - 30              &                &                        \\ 
          &  Nov 10   &   6.0$\pm$0.1        &  [3.8] - 30              &                &                        \\       
\hline
V2493 Cyg &  Nov 77   &   9.4$\pm$0.1        &   [3.5] - 3              &                &                        \\ 
          &  Aug 06   &   9.9$\pm$0.1        & IRAC[3.4] - 31           &  9.2$\pm$0.1   &  IRAC[4.5] - 31        \\        
\hline\hline

\end{tabular}}
\end{center}
\medskip
\scriptsize

- References to the Table: (1) Cohen 1974; (2) Elias 1978; (3) Cohen \& Kuhi 1979; (4) Rydgren 1984; (5) McCabe 2006; 
(6) Hartmann et al. 2005; (7) Rydgren et al. 1976; (8) Rydgren \& Vrba 1983; (9) Kenyon et al. 1994; (10) Thi et al. 2001;
(11) Rebull et al. 2006; (12) Breger et al. 1981; (13) Glass \& Penston 1974; (14) Appenzeller et al. 1983;
(15) Hughes et al. 1994; (16) Sipos et al. 2009; (17) Kun et al. 2011; (18) Cohen \& Schwartz 1983: (19) Liseau et al. 1992; 
(20) Aspin \& Sandell 1994; (21) Aspin \& Sandell 1997; (22) Chen et al. 2009; (23) Fischer et al. 2012; 
(24) Muzerolle et al. 2005; (25) Aspin 2011b; (26) Kenyon \& G\'{o}mez 2001; (27) Persi et al. 2007; (28) Hodapp et al. 1996;
(29) K\'{o}sp\'{a}l et al. 2007; (30) Aspin 2011a; (31) Guieu et al. 2009.\\
$^a$ This value refers to the total binary system, although the primary and the secondary have been resolved and separately observed.\\
$^b$ ISO-SWS continuum flux near H$_2$ lines.
\bigskip

\end{table}    


\begin{table}
\begin{center} 

\caption{N and Q band photometry of EXors. \label{variabNQ:tab}} 
\medskip
{\scriptsize
\begin{tabular}{l|c|cc|cc}
\hline
\medskip
Source    &  Epoch    &        \multicolumn{2}{c}{N band}               &       \multicolumn{2}{c}{Q band}         \\ 
          &           &  (Jy)          &  [$\lambda_{eff}$] - Ref       &    (Jy)        & [$\lambda_{eff}$] - Ref \\                   
\hline
UZ Tau E  &  73/74    &  1.6 $\pm$ 0.3  &  [10] -    1      &                 &                       \\
          &  Nov 76   &  1.3 $\pm$ 0.3  &  [10] -    2      &                 &                       \\
          &  Dec 76   &  1.0 $\pm$ 0.2  &  [10] -    2      &                 &                       \\
          &  Jan 81   &  1.1 $\pm$ 0.1  &  [10] -    3      & 0.8             &  [20] - 3             \\
          &  83       & 1.38 $\pm$ 0.03 &  IRAS[12]- 4      & 1.76 $\pm$ 0.04 &  IRAS[25] -  4        \\
          & Dec 92$^a$&  1.06 $\pm$ 0.04&  [9.5] -   5      &                 &                       \\
          & Nov 99$^b$&  1.5 $\pm$ 0.1  &  [10.8] -  6      &  1.6 $\pm$ 0.2  &  [18.1] -  6          \\
\hline
VY Tau    &  73/74    & $<$ 0.35        &  [10] -    1      &                 &                       \\
          &  Dec 73   &  0.3 $\pm$ 0.05 &  [10.8] -  7      &                 &                       \\
          &  83       & 0.20 $\pm$ 0.03 &  IRAS[12]- 4      & 0.29 $\pm$ 0.03 &  IRAS[25] -  4        \\
          &  Nov 99   & 0.07 $\pm$ 0.01 &  [10.8] -  6      &                 &                       \\
\hline       
DR Tau    &  72/74    &  2.2            &  [10] -    1      &                 &                       \\
          &  Nov 73   &  1.3            &  [11.1] -  7      &                 &                       \\
          &  83       & 3.16 $\pm$ 0.03 &  IRAS[12]- 4      & 4.30 $\pm$ 0.05 &  IRAS[25] -  4        \\ 
          &  96/98    & 2.4$^c$         &  [9.6] -   8      &                 &                       \\ 
          &  Dec 81   & 3.8             &  [10] -    3      & 2.9             &  [20] - 3             \\
          &  Dec 82   & 2.9             &  [10] -    3      &                 &                       \\
          &  96/98    & 2.4$^c$         &  [9.6] -   9      &                 &                       \\
          &  Dec 02   & 2.0             &  [11.9] -  10     &                 &                       \\
\hline
V1118 Ori &  Jan 02   &  $<$ 0.5        &  MSX[12] - 11     &                 &                       \\ 
          &  Jan 06   & 0.07 $\pm$ 0.01 &  [10.4] - 11      &                 &                       \\      
\hline
NY Ori    &  83       & 10 $\pm$ 3      &  IRAS[12]- 4      &                 &                       \\
\hline
V1143 Ori &  83       & 0.17 $\pm$ 0.04 &  IRAS[12]- 4      & 0.10 $\pm$ 0.06 &  IRAS[25] -  4        \\       
\hline
EX Lup    &  83       & 0.80 $\pm$ 0.03 &  IRAS[12]- 4      & 1.09 $\pm$ 0.03 &  IRAS[25] -  4        \\    
          &Feb/Sep 97 &(0.7-0.8)$\pm$0.1& ISOPHOT[12]- 12 & (0.9-1.3)$\pm$0.2 &  ISOPHOT[20] - 12     \\   
\hline

\end{tabular}}
\end{center}
\end{table} 
\addtocounter{table}{-1}

\begin{table}
\begin{center} 
\caption{Continued. \label{variabNQ:tab}} 
\medskip
{\scriptsize
\begin{tabular}{l|c|cc|cc}
\hline
\medskip
Source    &  Epoch    &        \multicolumn{2}{c}{N band}               &       \multicolumn{2}{c}{Q band}         \\ 
          &           &  (Jy)          & [$\lambda_{eff}$] - Ref        &    (Jy)       &  [$\lambda_{eff}$] - Ref \\                   
\hline
V1180 Cas &  83       &  $<$ 0.25       &  IRAS[12]-  13                & 0.72  FQUAL 3   &  IRAS[25] -  13       \\       
\hline
V512 Per  &  $<$ 83   &6.54$\pm$0.07    & [10.2] - 14                   & 25$\pm$1        &   [19] - 14           \\ 
          &  83       & 11.2 FQUAL3     & IRAS[12] - 13                 & 42.8  FQUAL 3   &  IRAS[25] - 13        \\
          &  Sep 04   &                 &                               & $>$ 21 (satur.) &  MIPS[24] - 15        \\
\hline
LDN1415IRS&  83       & 0.15$\pm$0.05   & IRAS[12]- 16                  & 0.49$\pm$0.06   &  IRAS[25] - 16        \\       
\hline
V2775 Ori &  Apr 05   &                 &                               & 0.68$\pm$0.03   &  MIPS[24] - 17        \\ 
          &  Nov 08   &                 &                               & 5.0$\pm$0.2     &  MIPS[24] - 17        \\      
\hline
V1647 Ori &  83       & 0.52            & IRAS[12]- 18                  & 1.2             &  IRAS[25] - 18        \\
          &  Mar 04   & 7$\pm$1         &  [11.2] - 19                  & 15.6            &  MIPS[24] - 20        \\ 
          &  Mar 07   & 0.23            &  [11.2] - 21                  & 0.44            &  [18.3] - 21          \\
          &  Sep 08   & 2.8$\pm$0.5     &  [11.2] - 19                  & 7$\pm$1         &  [18.3] - 19          \\             
\hline
GM Cha    &  Apr 04   & 1.15$\pm$0.09   &  [12.9] - 22                  &                 &                       \\        
\hline
OO Ser    &  Oct 95   &  6.4            &  [11.7] - 23                  &  12.5           &  [20.6] - 23          \\  
          &  Feb 96   &  4$\pm$1        &  [12] - 24                    &  29$\pm$3       &  [25] - 24            \\
          &  Apr 96   &  4.1$\pm$0.1    &  [12] - 24                    &  39$\pm$1       &  [25] - 24            \\
          &  Sep 96   &  3$\pm$1        &  [12] - 24                    &                 &                       \\
          &  Oct 96   &  4$\pm$1        &  [12] - 24                    &  21$\pm$2       &  [25] - 24            \\
          &  Mar 97   &  2.98$\pm$0.06  &  [12] - 24                    &  18$\pm$2       &  [25] - 24            \\
          &  Apr 97   &  3$\pm$1        &  [12] - 24                    &  19$\pm$2       &  [25] - 24            \\
          &  Sep 97   &  2.0$\pm$0.8    &  [12] - 24                    &  17$\pm$2       &  [25] - 24            \\
          &  Apr 04   &                 &                               &  13$\pm$2       &  MIPS[24] - 24        \\
          &  Oct 04   &  0.6$\pm$0.1    &  [10.4] - 24                  &                 &                       \\
\hline
V2492 Cyg &  83       &  3.4$\pm$0.3    & IRAS[12] - 25                 & 6.6$\pm$0.6     &  IRAS[25] - 25        \\
          &  Jul 96   &  2.5$\pm$0.1    & MSX[12.13] - 25               & 3.1$\pm$0.2     &  MSX[21.4] - 25       \\
          &  Jul 06   &                 &                               & 3.4$\pm$0.3     &  MIPS[24] - 25        \\      
\hline
V2493 Cyg &           &                 &                               &                 &                       \\        

\hline\hline

\end{tabular}}
\end{center}
\medskip
\scriptsize

- References to the Table: (1) Cohen 1974; (2) Elias 1978; (3) Rydgren et al. 1984; (4) Weaver \& Jones 1992 (5) Ghez et al. 1994; 
(6) McCabe 2006; (7) Rydgren et al. 1976; (8) Thi et al. 2001; (9) Thi et al. 2001; (10) Przygodda et al. 2003; 
(11) Lorenzetti et al. 2007; (12) Sipos et al. 2009; (13) IRAS Point Source Catalog; (14) Cohen \& Schwartz 1983; 
(15) Chen et al. 2009; (16) Stecklum et al 2007; (17) Fischer et al. 2012; (18) \'{A}brah\'{a}m et al. 2004; 
(19) Aspin et al. 2009; (20) Muzerolle et al. 2005; (21) Aspin et al. 2008; (22) Persi et al. 2007; (23) Hodapp et al. 1996; 
(24) K\'{o}sp\'{a}l et al. 2007; (25) Aspin 2011a; \\
$^a$ Spectrophotometry of the resolved component UZ Tau E in the 8-13 $\mu$m window: observations at close-by wavelengths are also available.\\
$^b$ These values refer to the total binary system, although the primary and the secondary have been resolved and separately observed.\\
$^c$ ISO-SWS continuum flux near H$_2$ lines.
\bigskip

\end{table} 

\clearpage   


\begin{table}
\begin{center} 
\caption{Short time-scale (days) mid-IR monitoring cases of a few EXors. \label{timescale:tab}} 
\medskip
{\scriptsize
\begin{tabular}{lccccc}
\hline
\medskip
Source          &  Epoch             &  Band    & N$_{obs}$  & $\mid x - \overline{x} \mid_{max}$   & 3$\sigma$  \\ 
                &                    &          &            &      \multicolumn{2}{c}{(mag)}                    \\
\hline
DR Tau          & Nov 87             &  L       &  13        &  0.16             & 0.03 \\
                & Sep 88$^a$         &  L       &  33        &  0.28             & 0.03 \\                
V1180 Cas       & Oct/Nov 09         &  L       &  12        &  0.17             & 0.03 \\
                & Oct/Nov 09         &  M       &  12        &  0.14             & 0.03 \\
V512 Per        & Sep 88             &  L       &   7        &  0.20             & 0.09 \\  
\hline\hline

\end{tabular}}
\end{center}
\medskip
$^a$ This sampling was obtained also on hours time-scale.
\medskip 
\end{table}    
\clearpage

\end{document}